\documentclass{epl}

\title{Coherent backscattering of ultrasound without a source}
\author{Eric LAROSE\inst{1,2} \and Oleg I. LOBKIS\inst{1} \and Richard L. WEAVER\inst{1}}

\institute{                    
  \inst{1}Department of Theoretical and Applied Mechanics, University of Illinois, Urbana, Illinois 61801.\\
    \inst{2}also at: Lab. de G\'eophysique Interne et Tectonophysique, BP53, 38041 Grenoble, France. Email: eric.larose@ujf-grenoble.fr
  }


\begin{document}

\maketitle

\begin{abstract}
Coherent backscattering is due to constructive interferences of 
reciprocal paths and leads to an enhancement of the intensity  of a multiply scattered 
field near its source. To observe this enhancement an array of 
receivers is conventionally placed close to the source. Our approach here is different. In a first experiment, we recover the coherent backscattering effect (CBE) within an array of sources and a distant receiver using time correlation of diffuse fields. The enhancement cone has an excellent spatial resolution.
 The dynamics of the enhancement factor is studied in a second experiment using correlation of thermal phonons at the same ultrasonic frequencies, without any active source. 

\end{abstract}

Coherent Backscattering Enhancement (CBE) is a consequence of interferences in multiple scattering or multiple reverberations of waves in a complex environment. It was first observed for electrons \cite{kawabata1982}, then in optics \cite{vanalbada1985,wolf1985
} where it exhibited as a doubling of the backscattered energy in the incident direction. Later, it was applied to various fields of wave physics, like ultrasound \cite{bayer1993,tourin1997,weaver2000a}, laser in cold atoms \cite{labeyrie1999}, and more recently to seismology \cite{larose2004a}. In these fields, CBE was shown to be an accurate way to measure transport mean-free paths of the medium, a quantity that characterizes its degree of heterogeneity. The origin of CBE is the constructive interference between long reciprocal paths in wave scattering or reverberation. This enhances the probability for the wave to return to the source by a factor of exactly two in open media, and three in closed cavities, which results in the local energy density enhancement by the same factor \cite{vantiggelen1998,akkermans2004
}. In a reverberant body, it was proved that the CBE factor increases from two to three with a dynamics driven by the Heisenberg time $T_H$ (also referred to as the break time of the cavity)~\cite{weaver2000a}. When the source and the receivers are placed inside the diffusive medium, the spatial extent of the cone is the order of the wavelength \cite{derosny2000,weaver2000a}. In all of these \textit{active} experiments, the observation of CBE requires that a receiver be placed close to the source. Unfortunately, the source-receiver configuration is sometime disadvantaging. First, it may break the reciprocity condition if the source and the receiver have not exactly the same orientation~\cite{vantiggelen2001a}. Second, placing a source close to the receiver is hard to perform as soon as the sizes of source and recording device are not negligible compared to the wavelength. In this paper, we propose that CBE can be
observed without a standard source-receiver configuration. The opportunity to retrieve the CBE passively will be provided by analyzing the time-correlation of diffuse fields generated by a set of distant deterministic sources (exp.~1) or diffuse field generated by thermal phonons (exp.~2).\\

The possibility of retrieving the impulse response (the Green function) of a complex medium by correlating records between two \textit{passive} sensors has been suggested several times in the literature \cite{rytov1989
}. Recently, Weaver and Lobkis \cite{lobkis2001
} proposed to cross correlate multiply scattered waves, a technique that intensively exploits the mesoscopic nature of diffuse wavefields, and has transpired to become a useful and accurate route to \textit{passive} imaging applications ranging from ultrasound and ocean acoustics to seismology \cite{derode2003a,roux2003,campillo2003a}.  
 In the following we apply this \textit{passive} technique to observe CBE \textit{within} an array of sources, the wave field being sensed by a distant receiver. By reciprocity, this configuration is analogous to an array of receivers and a distant source. \\

\begin{figure}[!htbp]
	\centering
		\includegraphics[width=8.6cm]{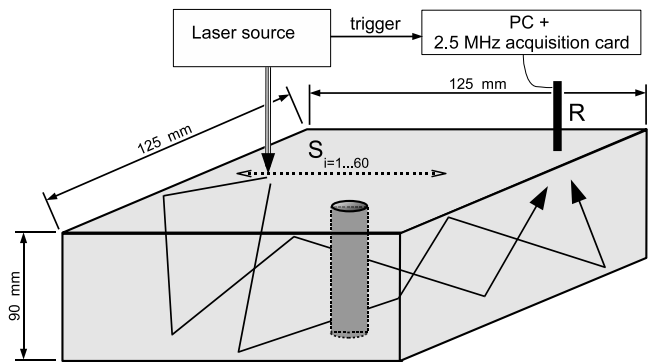}
	\caption{First experimental setup: Correlation of actively generated diffuse fields. A series of 60 laser shots $S_i$ is performed on the top of a reverberant aluminum body, the ultrasonic wave field is sensed by a distant receiver $R$. CBE is observed within the $S_i$'s.}
	\label{fig1}
\end{figure}

 The configuration of the first experiment (correlation of an actively generated diffuse field) is depicted in Fig.~\ref{fig1}. 
A piezoelectric transducer $R$ is oil-coupled to the top surface of an irregularly shaped object whith dimensions $\sim$90$\times$125$\times$125~mm. A hole is drilled through the block to break much of its symmetry and enhance the ray chaos. A Q-switched 60~mJ Nd-Yag infrared laser with pulse duration $\sim$8~ns and a beam diameter $\sim$1.5~mm excites elastic waves by means of ablation mechanism at position $S_i$, resulting waves then reverberate thousands of times in the aluminum body. The receiver is far (several wavelengths) from the array of sources, its precise position does not play any role in our experiment.  The absorption time $T_a$ varies with frequency and is the order of  tens of milliseconds, so the records are well above the noise level over the whole $T$=210~ms acquisition duration. The signal $\varphi_i(t)$ for the source $i$ is then low-pass filtered, digitized and passed to the computer for processing. The laser source triggers the acquisition. The laser is mounted on a scanning motor of 1.5~mm step, the pitch-catch sequence is repeated over $i=1..60$ positions. When the last impulse response is obtained, the laser is moved back to the initial position. To check the change of the medium due to temperature, one additional response $\varphi_1'(t)$ is acquired and compared to the initial signal $\varphi_1(t)$. Within the few minutes of the experiments, the thermal variations \cite{lobkis2003} were found to have no effect on the results presented in this article. The time-averaged correlation is processed for each couple of sources $ij$ for times $\tau$ ranging from 0 to 60~ms:

\begin{equation}
C_{ij}(\tau)=\int_{0}^{T} \varphi_i(t) \varphi_j(t+\tau)dt.
\end{equation}

\begin{figure}[!htbp]
	\centering
		\includegraphics[width=8cm]{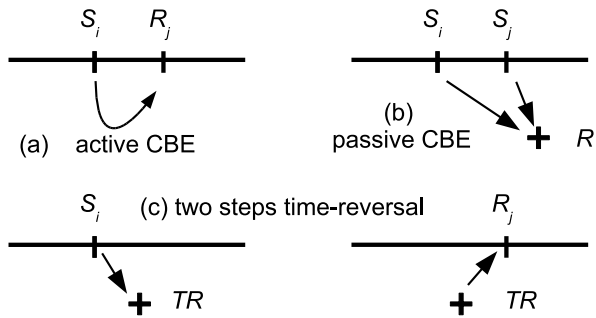}
	\caption{Three possible experimental configurations to observe CBE. (a) Conventional setup: $S_i$ is a source, the wave fields sensed in $R_j$ after propagating in the medium. (b) \textit{Passive} CBE: $S_i$ and $S_j$ are two sources, the wave fields are sensed in $R$ before correlation. (c) Time-reversal experiment: the field is emitted in $S_i$ and sensed in $TR$, then time-reversed and send back to the medium to be probed in $R_j$.}
	\label{fig1b}
\end{figure}

 The configuration and processing of this first experiment are analogous to a one-channel time reversal experiment \cite{draeger1999a}, where the field is emitted in $S_i$, sensed and time-reversed in $TR$, back propagates in the medium and is finally received in $S_j$. The analogy between time-reversal and correlations has been developed by Derode \textsl{et al.}\cite{derode2003a} and holds as long as reciprocity is valid. A schematic view of the analogy is presented in Fig.~\ref{fig1b}. The link between time-reversal and CBE was also developed by De Rosny \textit{et al.} \cite{derosny2005}. In the open scattering medium they used, they found an energy enhancement of the time-reversed field if the source $S$ and the receiver $TR$ (the time-reversal device) are close together. As we will see, an enhancement is also  visible in a finite body around $S$ even if  $R$ and $S$ are far apart.\\

 The correlations $C_{ij}$ are filtered with a $\Delta f/f=$50\% bandwidth, the intensity is averaged $\left\langle \right\rangle$ over all the $ij$ couples available for a given distance $r=S_i-S_j$, and additionally over times 30~ms$<\tau<$60~ms greater than the Heisenberg time of the cavity, which at 87~kHz is $T_H\approx 16$~ms. The spatial enhancement of CBE is defined as the following ratio:

\begin{equation}\label{exp}
CBE_{spatial}(r)=\frac{\left\langle C_{ij}^2(r) \right\rangle}{\left\langle C_{ij}^2(r\gg \lambda) \right\rangle}.
\end{equation}

 If the integration time $T$ and the correlation time $\tau$ are well above the Heisenberg time, then a decomposition of the wave field in uncorrelated modes $u_n$ is relevant and the ensemble averaged correlations reads\cite{weaver1994}:
\begin{equation}\label{modes}
\left \langle C_{ij}^2\right\rangle=\left\langle\sum_n u_n(S_i)^2u_n(S_j)^2u_n(R)^4\right\rangle
\end{equation}
We assume that the receiver $R$ is far from the array of sources $S_{i-j}$, and that the modes are random gaussian variables with a spatial correlation of: $\left\langle u_n(S_i)u_n(S_j)  \right\rangle=J_0(\frac{2\pi f}{c}r)$ where $J_0$ is the Bessel function of zero order. Therefore, after an expansion of Eq.~\ref{modes}, all the terms $\left\langle u_n(R)u_n(S_{iorj})  \right\rangle=0$ vanish. Each mode having the same mean-intensity $\left\langle u^2\right\rangle$, Eq.~\ref{modes} rewrites:

\begin{equation}
\left \langle C_{ij}^2\right\rangle=\left\langle u^2\right\rangle^2\left( \left\langle u^2\right\rangle^2 +2\left\langle u_n(S_i)u_n(S_j)  \right\rangle^2          \right)
\end{equation}

 Therefore the normalized CBE at a given frequency $f$ averaged over the $\Delta f$ frequency band takes the theoretical form:
\begin{equation}\label{theo}
\frac{1}{\Delta f}\int_{f-\frac{\Delta f}{2}}^{f+\frac{\Delta f}{2}} 1+2J_0^2\left(\frac{2\pi f}{c}r\right)df.
\end{equation}
\begin{figure}[!hbp]
	\centering
		\includegraphics[width=8.6cm]{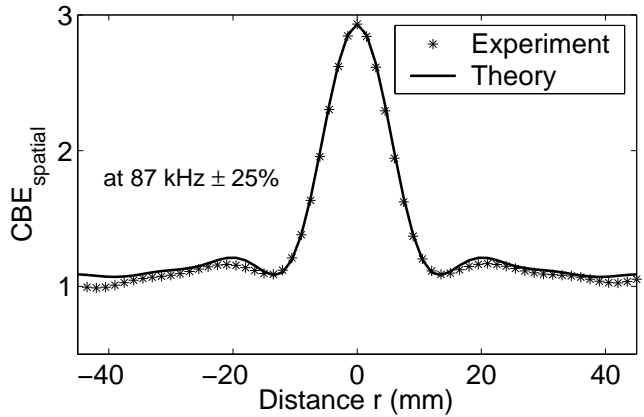}
	\caption{Spatial extent of the coherent backscattering effect for time $\tau$ greater than the Heisenberg time. Experimental data are processed following Eq.~(2) after a $f$=87~kHz 50\% bandwidth filter. Theory comes from Eq.~(4).}
	\label{fig2}
\end{figure}

Theoretical and experimental CBE are displayed in fig.~\ref{fig2} around $f$=87~kHz.  The relevant velocity here is the Rayleigh wave speed $c=2.9$~mm/$\mu$s. The theoretical curve satisfactorily fits the experimental data, including the secondary lobes of the Bessel function. The \textit{passive} setup allowed an optimal spatial resolution, which was only limited by the size of the source, and could be well bellow 1~mm if we chose to use a focusing optical lens. In an \textit{active} source-sensor experiment, CBE is observed under the condition that sources and receivers have the same symmetry. This condition is naturally ensured here since we evaluate the impulse responses between reproducible sources $S$.\\

Let's now focus on the dynamics of the enhancement factor. In an \textit{active} experiment the enhancement factor was found to vary with time $\tau$ in the coda \cite{weaver1994,weaver2000a} as $\label{eq5} 3-b(\tau/T_H)$, where $b$ is the Fourier transform of the  Dyson's two-level cluster function. $b$ decreases from one at short times to zero at late times. Therefore the enhancement factor should increase from two for $\tau\ll T_H$ to three for $\tau\gg T_H$.\\

Because correlations $C_{ij}(\tau)$ are not exactly the impulse response between $i$ and $j$ but also have residual fluctuations resulting from incomplete ensemble or time averages, the enhancement factor and its dynamics are here slightly different from that of an \textit{active} CBE. The intensity $N(T)$ of these fluctuations  of the diffuse field correlations has been studied\cite{weaver2005c}, but this theory does not hold for $S_i=S_j$; a theory at backscatter is needed. Fluctuations of the correlations of thermal noise do not depend on the distance between the receivers but only on the record length $T$. Therefore, instead of studying the dynamics of CBE with an actively generated diffuse field, we chose to realize another experiment based on correlation of thermal noise\cite{weaver2001}. \\

\begin{figure}[!htbp]
	\centering
		\includegraphics[width=8.6cm]{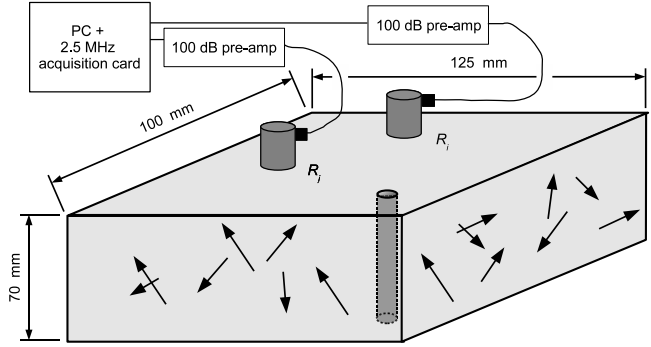}
	\caption{Second experiment: correlation of thermal noise. Two acoustic emission transducers sense the ultrasonic thermal noise at $R_i$ and $R_j$. Waveforms are acquired after a 100~dB amplification, digitized and stored on a computer. }
	\label{fig4}
\end{figure}

The second experimental setup is depicted in Fig.~\ref{fig4}. Because thermal noise is much weaker than the diffuse field obtained in the first experiment, we need to employ more sensitive acoustic  emission transducers, noted $R_i$ and $R_j$, along with two 100~dB pre-amplifiers (from Panametrics). The distance $R_i-R_j$ is much greater than one wavelength. The aluminum block has a size comparable to that of the first experiment. Thermal noise is recorded during a period of time T=400~s at a sampling frequency of 2.5~MHz; 6 series of acquisitions at different position pairs $i\neq j$ are performed. The total amount of waveforms collected represents 24~Gbytes of data. 
 The useful thermal noise from the aluminum block represents only 40\% of the record, but the additional electronic noise is white and $\delta$-correlated, and therefore average down in the correlation. Correlations are then processed, and the dynamics (time-dependence) of the enhancement factor is evaluated through the following ratio, which compensates for the transducers' sensitivities\cite{weaver2003b}:

\begin{equation}\label{eq6}
CBE_{dynamic}(\tau)=\sqrt{\frac{C_{ii}^2(\tau)C_{jj}^2(\tau)}{C_{ij}^2(\tau)C_{ji}^2(\tau)}}
\end{equation}

\begin{figure}[!htbp]
	\centering
		\includegraphics[width=8.6cm]{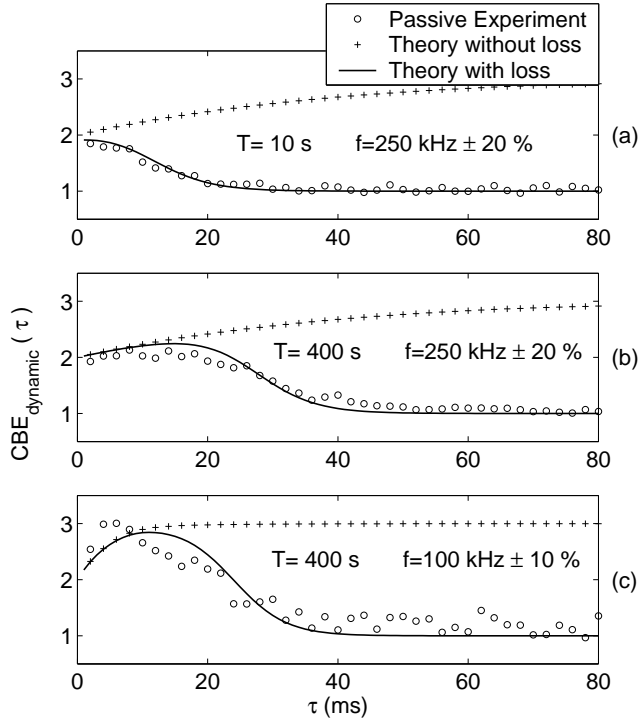}
	\caption{Dynamics of CBE obtained by correlating thermal ultrasonic noise. Experimental data are processed following Eq.~\ref{eq6}. Theory without loss (absorption) is $3-b(\tau)$~\cite{weaver1994}, and theory with loss is from Eq.~\ref{eq7} \cite{note}. }
	\label{fig5}
\end{figure}

We implicitly assume a spatial averaging over the six available positions, and a time-averaging  in a 2~ms sliding window centered around $\tau$. Results are presented in Fig.~\ref{fig5} for two different frequencies (100 and 250~kHz) and two different record lengths ($T=10$~s and $T=400$~s). The Heisenberg time of the block around 100~kHz is $T_H=18$~ms, and is $T_H=53$~ms around 250~kHz. The fluctuations $N(T)$, which are mainly determined by the amplifiers electronic noise, show a constant intensity in time $\tau$. They also linearly increase with the record length $T$ whereas the intensity of the Green function in the correlations increases quadratically. Because of the large surface of contact of the transducers  and the oil couplant, the absorption time describing the intensity decay is much smaller than in the first experiment: $T_a=4.5$~ms around 100~kHz and $T_a= 5$~ms around 250~kHz. Therefore the mean intensity of the correlation is well described by $E_0e^{-\tau/T_a}+N(T)$. At early time $\tau$, fluctuations are negligible: $E_0/N(T)=400$ at $\tau=1$~ms and for $T=400$~s. As $\tau$ increases, correlation intensity decreases, and fluctuations are finally dominating the correlations for times greater than 50~ms. We propose the following as a model for the \textit{passive} CBE dynamics:

\begin{equation}\label{eq7}
CBE_{dynamic}(\tau)=\frac{E_0\left(3-b(\tau/T_H)\right)e^{-\tau/T_a}+N(T)}{E_0e^{-\tau/T_a}+N(T)}
\end{equation}
This theory  takes into account loss in the elastic body $T_a$ and fluctuations in the correlations $N(T)$. It is displayed in solid line in Fig.~\ref{fig5} and is found to well fit the experimental data (circles): for early $\tau$, CBE is increasing from two to three until fluctuations dominate and the enhancement factor drops to one.\\

In this paper, we conducted two experiments on ultrasound propagating in an elastic body. In the first one, we retrieved the spatial shape of the coherent backscattering enhancement by correlating diffuse fields. This \textit{passive} technique allows to observe CBE within an array of sources using only one or two distant receivers, and the reciprocity principle. Additionally, CBE shows a spatial resolution much greater than previous \textit{active} experiments \cite{weaver2000a,derosny2000,larose2004a}: the resolution here is no longer limited by the source-sensor distance but only by the distance between two consecutive sources. We also recalled the link between time-reversal, correlations and CBE. Godin~\cite{godin2006} recently demonstrated that in a non-reciprocal medium, correlations still yield the Green function. Therefore, as observed in an \textsl{active} experiment~\cite{derosny2005}, we suspect that breaking the reciprocity will destroy our \textsl{passive} CBE. In the second experiment presented here, we correlated ultrasonic noise made of thermal phonons. The dynamics of CBE was studied and the main features were recovered: the increase of the enhancement factor from two to three is driven by the Heisenberg time. The effect of loss in the aluminum body and fluctuations in the correlations were also evaluated and proved to be a possible limitation.


\acknowledgments
This work was funded by the National Science Foundation EAR 0543328. The first author wish to thank J. De Rosny, 
M. Campillo and B. Van Tiggelen for enlightening discussions and 
for support, and acknowledges a Lavoisier grant from the French Ministry of Foreign Affairs.



\begin{thebibliography}{0}
\bibitem{note}
At low frequency and late times ($\tau>30~\mu$s), the enhancement factor is slightly greater than one. This is due to the noisy response of one transducer in the narrow frequency band investigated.
\end{thebibliography}

\end{document}